\documentclass{moriond}

\def\be{\begin{equation}}
\def\ee{\end{equation}}
\def\bea{\begin{eqnarray}}
\def\eea{\end{eqnarray}}

\newcommand{\Photo}{\includegraphics[height=35mm]{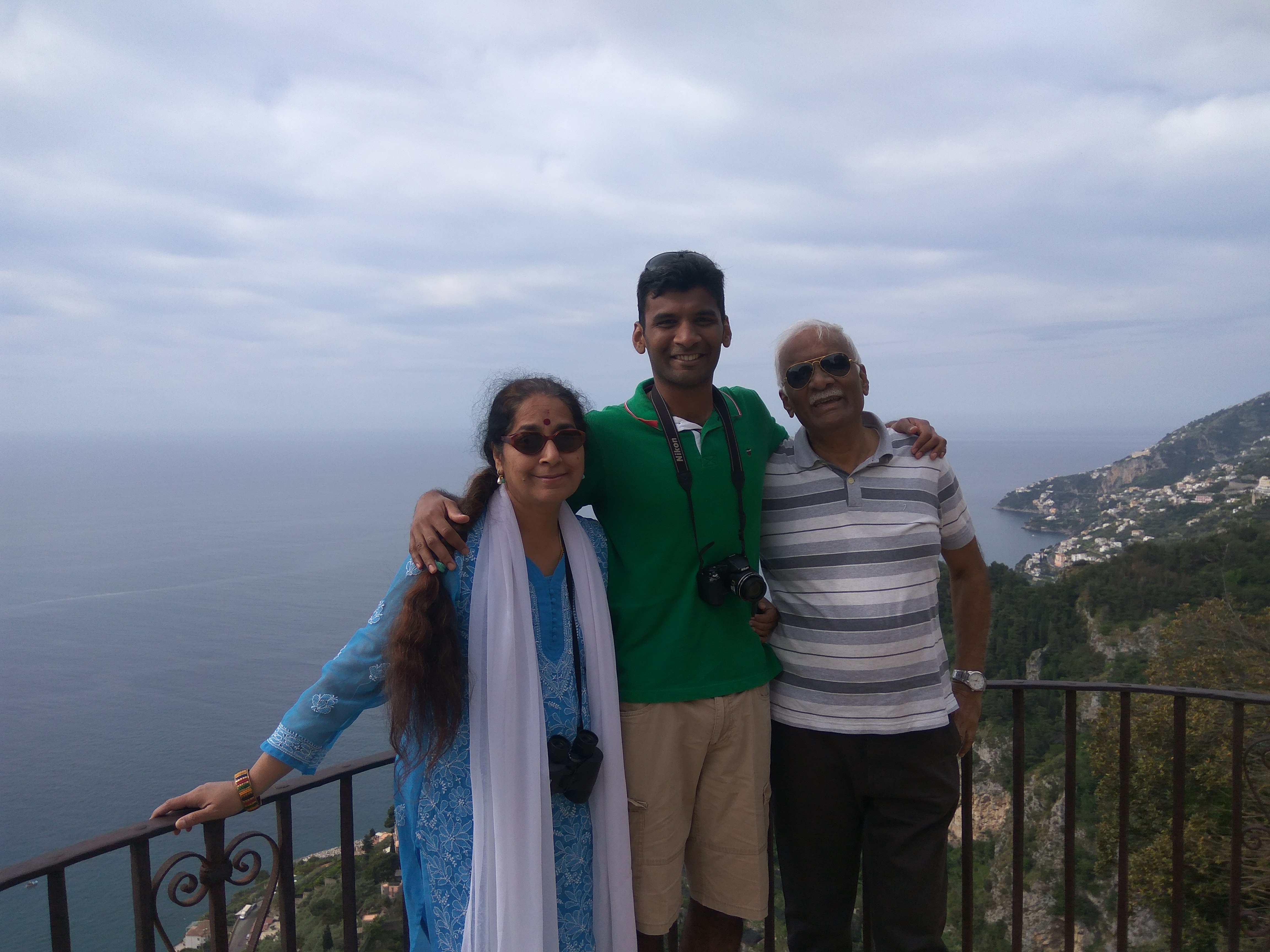}}

\begin{document}
\vspace*{4cm}
\title{$B$ anomalies: From warped models to colliders}

\author{ Abhishek M. Iyer }

\address{INFN Sezione di Napoli,\\
via Cintia, 80126 Napoli Italia}

\maketitle\abstracts{
We address the  anomalies in the semi leptonic decays of $B$ mesons  in a warped custodial framework. Two possible solutions of lepton non-universality are discussed: A) The muon singlets couple non-universally to NP and B) The non-universality is in the coupling of lepton doublets. Both these scenarios are characterized by different predictions for rare Kaon decays. An essential feature of these scenarios is that the electron contribution to the Wilson coefficients (WC) is non-vanishing, thereby offering possibilities for different patterns of solutions. Beginning with a generic $Z'$ model, we demonstrate how the observation of the ratio $N_{\mu\mu}/N_{ee}$ can be mapped to a given pattern of WC which  satisfy the anomalies.}

\section{Introduction}

The observation of lepton flavour universality (LFU) violation in the semi-leptonic decay of $B$ mesons has fueled several  new physics (NP) explanations. In the neutral current sector ($b\rightarrow sll $ transitions), these observations are parametrized through the measurement of the following ratio \cite{Aaij:2014ora}:
\begin{eqnarray}
	R_K=\left.\frac{\mathcal{B}(B^+\rightarrow K^+\mu^+\mu^-)}{\mathcal{B}(B^+\rightarrow K^+e^+e^-)}\right\vert_{q^2=1-6~GeV^2}=0.745^{+0.090}_{-0.074}~(stat)\pm 0.036~(syst)\nonumber\\
\end{eqnarray}
The SM predicts this ratio to be $R_K^{SM}=1.0003\pm0.0001$,  implying a $\sim 2.6~\sigma$ deviation from the SM expectation. Furthermore, this observation was corroborated by the measurement of $R_{K^*}=\frac{\mathcal{B}(B^0\rightarrow K^{*0}\mu^+\mu^-)}{\mathcal{B}(B^0\rightarrow K^{*0}e^+e^-)}$,
 signaling a  $2.4\sigma$ deviation for low $q^2$ and $\sim 2.5~\sigma$ for medium $q^2$ bins.

In this talk, we will consider their explanation in a Randall Sundrum model with custodial symmetry \cite{Randall:1999ee}. This model is characterized by the presence of  fermions and gauge bosons in the bulk. On account of this construction, the warped framework admits non-universal coupling of the SM fermions to the KK gauge bosons, thereby offering an explanation to the observed anomalies in the $b\rightarrow sll$ sector.
Two scenarios of the solutions will be discussed:A) Unorthodox case where non-universality exists in the coupling of lepton singlets to NP and B) Standard scenario where the lepton doublets couple non-universally to NP \cite{DAmbrosio:2017wis}. 
An important aspect of this framework is that the electron coupling to NP is non-vanishing. This results in a possibility where a multi-dimensional (2 or 4) fit to data, (involving both electrons and muons) is possible. With this as a background, we devise a strategy  where the information on the extent of electron and muon contribution can be extracted at high $p_T$ hadron experiments. This technique is general for any model with heavy neutral vector bosons. We demonstrate that any differences in the number of di-electrons and di-muons must be due NP effects and not due to experimental uncertainties in their detection.

\section{$b\rightarrow sll$ anomalies}\label{subsec:prod}
\begin{figure}
	\begin{center}
		\includegraphics[width=5.2cm,height=9cm,keepaspectratio]{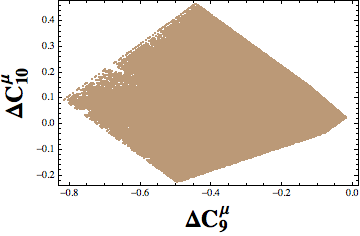} \includegraphics[width=5.2cm,height=9cm,keepaspectratio]{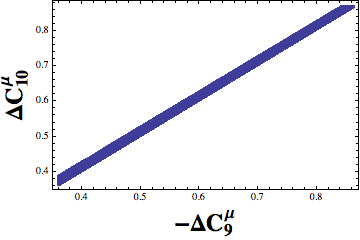}
	\end{center}
	\caption{ Scenario \textbf{A}: Plots gives the correlation in the $C_9$ and $C_{10}$ parameter plane for both the electron and the muon. 
		We use $M_{KK}=3~TeV$}
	\protect\label{hypothesis1}
\end{figure}

The geometry of the RS model, facilitates differential localization of the fermion fields in the bulk by appropriate choice of bulk mass parameters $c$. This offers an understanding of the hierarchical pattern of the Yukawa couplings.
Consequently, the couplings of the fermions to the KK gauge bosons are different thereby leading to the possibility of FCNC at tree level.
For the quarks we assume the first two generation couple universally, implying the presence of a $U(2)$ symmetry while the non-universality is due to the third generation couplings. Additionally, we assume that the down type singlets also couple universally. The bulk mass parameters for the third generation quark is chosen to lie in the range $c_{Q_3}\in[0.4,0.5]$ and $c_{t_R}\in[0,0.5]$ while the ranges for other fields are chosen to be $c>0.5$.  These numbers, in particular for the down type quarks, are motivated keeping in mind the constraints from $\Delta F =2$ processes.
 For the lepton sector we consider the following two scenarios:\\
\textbf{Scenario A:} The non-universality is due the coupling of the muon singlets while the doublets couple universally to NP.\\
\textbf{Scenario B:} In this case the doublets couple non-universally to NP while the coupling of the singlets is assumed to be universal.\\
In order to describe the observed deviations, we evaluate the NP effects to the Wilson coefficients of the following operators:
\begin{eqnarray}
\mathcal{L}\supset \frac{V^*_{tb}V_{ts}G_F\alpha}{\sqrt{2}\pi}\sum_i C_i\mathcal{O}_i
\label{effectivelagrangian}
\end{eqnarray}
In warped custodial models, there are four additional contributions to tree-level FCNC: $X\in {Z_{SM,Z_X,Z_H,\gamma^{(1)}}}$.
The effective four fermion interactions describing the  $b\rightarrow sll$ transitions in this case can be written as:
\begin{equation}
\mathcal{L}_{NP}\subset \sum_{X=Z_{SM},Z_H,Z_X,\gamma^{(1)}}X_{\mu}\left[\alpha^{bs}_L(X)(\bar s_L\gamma^\mu b_L)+\alpha^{bs}_R(X)(\bar s_R\gamma^\mu b_R)+\bar l\left(\alpha^l_V(X)\gamma^\mu-\alpha^{l}_{A}(X)\gamma^\mu\gamma^5 \right)l \right]
\label{rseff}
\end{equation}
where $\alpha^l_{V,A}(X)=\frac{\alpha^l_L(X)\pm\alpha^l_R(X)}{2}$. 
Matching Eq.\ref{rseff} to the operators in  Eq. \ref{effectivelagrangian}, the Wilson co-efficients for each gauge field $X$ is given as:
\begin{eqnarray}
\Delta C_9&=&-\frac{\sqrt{2} \pi}{M_X^2 G_F\alpha}\alpha^{bs}_L(X) \alpha^l_{V}(X),\;\;\;\;\;\;\Delta C'_9=-\frac{\sqrt{2} \pi}{M_X^2 G_F\alpha}\alpha^{bs}_R(X) \alpha^l_{V}(X)\nonumber\\\Delta C_{10}&=&\frac{\sqrt{2} \pi}{M_X^2 G_F\alpha}\alpha^{bs}_L(X) \alpha^l_{A}(X),\;\;\;\Delta C'_{10}=\frac{\sqrt{2} \pi}{M_X^2 G_F\alpha}\alpha^{bs}_R(X) \alpha^l_{A}(X)
\label{wc}
\end{eqnarray}
We work in the mass basis of the up-type quarks such that the rotation matrix $D$ in the down sector corresponds to the $V_{CKM}$. Under the assumption that the down type quark singlets couple universally to NP, $C'_{9,10}$ is zero. We now evaluate the contribution to $C_{9,10}$ for the two scenarios discussed above:\\
\textbf{Scenario A}:The ranges chosen for $c$ parameter scan for  the leptons are $c_{\mu_L}=c_L\in[0.45,0.55]$ and $c_{\mu_R}\in[0.45,0.55]$. Fig. \ref{hypothesis1} gives the results of the scan.
The scan in this scenario has some interesting features: As shown in the left plot of Fig \ref{hypothesis1}, we find that there exists region in the parameter space where $\Delta C^\mu_{10}$ is vanishing while $\Delta C^\mu_{9}$ is considerable. This is an artefact of the similar scanning ranges for the doublets and muon singlets and corresponds to the case where $\alpha^l_L(X)=\alpha^l_R(X)$. In such a case,  there exist a parameter space where the corresponding electron contributions are also small, thereby reducing the fit to a 1-D case. However, the fits can be obtained just outside the $2~\sigma$ region. This tight constraint can be relaxed by involving the electrons, thereby making it a 4-D fit to the data.\\
\textbf{Scenario B:} This is the standard scenario, where the electron contributions are typically small. The non-universality in the lepton sector is due to the differential coupling of the muon doublets to NP. Additionally, we allow the coupling of the  third generation lepton doublet to also be a free parameter (subject to fitting the tau mass). 
Right plots of Fig. \ref{hypothesis1} gives the results of the analysis  illustrating the correlation in the $C^\mu_{10}-C^\mu_{9}$ plane. The 2-D fit can be reduced to a 1-D fit  when $C^\mu_{10}=-C^\mu_{9}$. 
\section{Correlating the pattern of solutions to high $p_T$ searches at colliders}
The different pattern of solutions to the $B$ anomalies, involving  electrons and/or muons can be probed at high energy $p$-$p$ collisions.  To begin with, we consider the following neutral current Lagrangian:
\begin{eqnarray}
\mathcal{L}_{eff}&=&\frac{\lambda_{bs}}{M^2}(\bar s\gamma_\mu b)\left[\lambda_e(\bar{e}\gamma^\mu e)+\lambda_\mu(\bar{\mu}\gamma^\mu\mu)+\lambda_\tau(\bar{\tau}\gamma^\mu \tau)\right]
+\ldots
\label{structure1}
\end{eqnarray}
Given the lagrangian in Eq. \ref{structure1}, the Wilson-coefficients for the $R(K^*)$ anomalies is given as
\begin{eqnarray}
C^e_{9}=-\frac{\sqrt{2}\pi}{G_F\alpha}\frac{\lambda_{bs}\lambda^V_e}{M^2}\;\;\;\;C^\mu_{9}=-\frac{\sqrt{2}\pi}{G_F\alpha}\frac{\lambda_{bs}\lambda^{V}_\mu}{M^2}\;\;\;
C^e_{10}=-\frac{\sqrt{2}\pi}{G_F\alpha}\frac{\lambda_{bs}\lambda^{AV}_e}{M^2}\;\;\;\;C^\mu_{10}=-\frac{\sqrt{2}\pi}{G_F\alpha}\frac{\lambda_{bs}\lambda^{AV}_\mu}{M^2}
\end{eqnarray} 
where $\lambda^{V}=\frac{L+R}{2}$ and $\lambda^{AV}=\frac{L-R}{2}$. $L(R)$ is the coupling strength of doublets (singlets) to $Z'$. As an illustration, we assume $L=R$, implying the axial-vector couplings of the leptons to $Z'$ vanish. In this case the ratio of $C_9$ for muons and electrons is simply
$\frac{C^9_\mu}{C^9_e}=\frac{\lambda^V_\mu}{\lambda^V_e}$. Notice here  that the quark dependence cancels out. 

Now consider the production of $Z'$ in $pp$ collisions with its corresponding decays into electrons and muons.
 The ratio of number di-muons to di-electrons in the $Z'$ mass bin at luminosity $\mathcal{L}$ is simply:
 $\delta=\frac{\sigma_{Z'}(\lambda^V_\mu)^2\mathcal{L}\epsilon_\mu}{\sigma_{Z'}(\lambda^V_e)^2\mathcal{L}\epsilon_e} =\frac{N_\mu}{N_e}$.
 As shown in Table. \ref{tab2}, electrons and muons are accompanied by different acceptance efficiencies. 
 
 \begin{table}[htb!]
 	\begin{center}
 		\begin{tabular}{ |c|c | c |  }
 			\hline
 			&$\epsilon_\mu$&$\epsilon_e$\\
 			\hline			
 			Simple Isolation($>1$ leptons)&59.33 & 39.79  \\
 			Mass cuts ($> 1000 GeV$)&58.79 & 39.61\\
 			\hline
 		\end{tabular}
 	\end{center}
 	\caption{Comparison of acceptance efficiencies   for electrons and muons for $m_{Z'}=3000$ GeV} 
 	\label{tab2}
 \end{table}
 In order to ensure that the measurement of this ratio is purely due to the difference in couplings of the $Z'$ to the leptons, the acceptance efficiencies must be made approximately similar \textit{i.e.} $\epsilon_e\simeq\epsilon_\mu$. This can be achieved by using the electron jet techniques and using the following substructure variables to identify the electron \cite{new}:\\
 A) Hadronic energy fraction: The jets corresponding to the electrons must be associated with very small $H-cal$ deposits. This helps in limiting QCD to a great extent and 
 B) Tracks: The leading jet is associated with exactly one track, while the subleading jet may have 0 or 1 track. The latter condition is useful in accepting events which would otherwise have been rejected by the standard electron identification criteria. Table \ref{tab2} gives a comparison of the efficiencies when electron jets techniques are employed and we find that the acceptance efficiencies are within $5-10\%$ of each other. The continuum SM background has been estimated and is similar for both the electrons and the muons.
 As a result,  any discrepancy in the observation of the ratio can be attributed to the difference in the couplings of the NP to electrons and muons. 
 As shown in the third column of Table \ref{tab3}, this technique can also be extended to tau jets and can be used to extract information on $C^\tau_9$, albeit with lesser accuracy than the leptons. This will have implications for processes like $B\rightarrow K^*\tau\tau$
\begin{table}
	\begin{center}
		\begin{tabular}{ |c|c | c | c| }
			\hline
			$m_{Z'}$ (GeV)	&$\epsilon_\mu$&$\epsilon_e$ (Electron jets)&$\epsilon_\tau$ (tau jets)\\
			\hline			
			2000&71.45 & 64.75&31.25  \\
			2500&66.35 & 63.06&37.28\\
			3000&58.79&60.37&40.88\\
			3500&51.68&59.50&43.98\\
			\hline  
		\end{tabular}
	\end{center}
	\caption{Comparison of efficiencies of electron jets and muons for different $m_{Z'}$ masses. A selectrion criteria of the $m_{\mu\mu}$ or $m_{j_0j_1}$ $(>1000)$ GeV is imposed. For the electron jets the QCD fake rate is $<1$ in $3\times 10^5$ events.
		For $\tau$ jets the QCD fake rate is $0.2\%$} 
	\label{tab3}
\end{table}

\section{Conclusions}
We discussed a   warped model with custodial symmetry to address the $B$ anomalies. The pattern of the solutions included contributions to the Wilson coefficients of both the electron and the muon. This provided a natural motivation to look at high $p_T$ experiments which can possibly probe the nature of these solutions. We construct a simple variable $N_{\mu\mu}/N_{ee}$ and demonstrate that this ratio can be mapped to the pattern of WC which  satisfy the anomalies

\section*{Acknowledgments}

AI were supported in part by MIUR under Project No. 2015P5SBHT and by the INFN research initiative ENP. A.I would also like to thank the organisers of Rencontres de Moriond  for the opportunity to present the talk.

\end{document}